\documentclass[aps,superscriptaddress,preprint]{revtex4}
\usepackage{url}
\usepackage{graphicx}
\usepackage{bm}
\begin{document}
%\doi{}
 %\issn{}
 %\jvol{00} \jnum{00} \jyear{2012}

%\markboth{Taylor \& Francis and I.T. Consultant}{Journal of Turbulence}

%\articletype{}

\title{Geotropic tracers in turbulent flows: a proxy for fluid acceleration}

\author{G. Boffetta,$^{\rm a}$ 
M. Cencini,$^{\rm b}$
\vspace{6pt} F. De Lillo,$^{\rm a}$
\vspace{6pt} and F. Santamaria$^{\rm a}$\\\vspace{6pt}
$^{\rm a}${\em{Dip. di Fisica and INFN, Universit\`a di Torino, 
via P.Giuria 1, 10125 Torino, Italy}};
$^{\rm b}${\em{Istituto dei Sistemi Complessi,  CNR, via dei Taurini 19, 00185
  Rome, Italy}}\\
\vspace{6pt}}
%\received{} }

\begin{abstract}

We investigate the statistics of orientation of small, neutrally buoyant, spherical tracers whose center of mass is displaced from the geometrical center. If
appropriate-sized particles are considered, a linear relation can be derived
between the horizontal components of the orientation vector and the same
components of acceleration. Direct numerical simulations are carried out,
showing that such relation can be used to reconstruct the statistics of
acceleration fluctuations up to the order of the gravitational acceleration.
Based on such results, we suggest a novel method for the local experimental
measurement of accelerations in turbulent flows.

\end{abstract}

\maketitle
%%%%%%%%%%%%%%%%%%%%%%%%%%%%%%%%%%%%%%%%%%%%%%%%%%%%%%%%%%%%%%%%%%%%%%
\section{Introduction}
\label{sec1}

The Lagrangian investigation of turbulence has dramatically
improved in the last few years in experimental techniques,
theoretical models and numerical simulations \cite{Toschi2009}. These
progresses benefited from the increased range of Reynolds numbers
accessible for investigation (in particular, for simulations) and the
improved accuracy of measurement techniques. On the theoretical side,
we have now phenomenological models able to quantitatively explain the
Lagrangian properties of turbulence such as, e.g.
the statistics of velocity increments \cite{Arneodo2008} and
accelerations \cite{Biferale2004}.
Grounded on the successes of Lagrangian investigations, recent
experimental and numerical studies started to investigate the motion
of complex objects in turbulent flows
\cite{Zimmermann2011,Zimmermann2011a,Zimmermann2012,Klein2013,Vincenzi2012,Parsa2012}. The motivations are both
fundamental and applicative. 
In this short note we suggest a possible technique to measure
turbulent accelerations without the need of particle tracking, by
means of the local measurement of the orientation of finite-size particles. The idea relies on
spherical particles whose average density is that of the carrier fluid (so
that they are neutrally buoyant), but whose center of mass is
displaced with respect to the geometrical center (implying that the
orientation is determined by the gravitational torque and that due to
the fluid). By means of direct numerical simulations (DNS) we show that information on particle orientation can be used to estimate fluid accelerations up to the order of gravitational acceleration. 

The paper is organized as follows. In Sect.~\ref{sec2} we discuss the
theoretical basis of the technique in its simplest
implementation. Sect.~\ref{sec3} presents some preliminary validation
of the method based on numerical simulations. Finally,
Sect.~\ref{sec4} is devoted to conclusions and perspectives.

%%%%%%%%%%%%%%%%%%%%%%%%%%%%%%%%%%%%%%%%%%%%%%%%%%%%%%%%%%%%%%%%%%%%%%
\section{The motion of  geotropic tracers}
\label{sec2}

%------------------------------------------------------------------------
\begin{figure}[htb!]
\centering
\includegraphics[width=5cm]{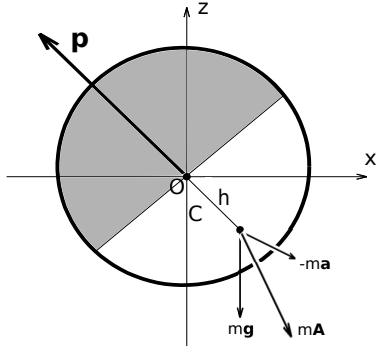}
\caption{An example of geotropic particle with black and white
  pattern. 
\label{fig1}}
\end{figure}
%------------------------------------------------------------------------

We consider the trajectory, $\bm x(t)$, of a neutrally-buoyant sphere 
small enough such that  its dynamics can be approximated by that of a passive tracer,
\begin{equation}
{d {\bm x} \over d t} = {\bm u}({\bm x},t)\,,
\label{eq:1}
\end{equation}
transported by a flow ${\bm u}({\bm x},t)$.  As sketched in
Fig.\ref{fig1}, we assume that the particle center of mass $C$ is
displaced by a distance $h$ with respect to the geometrical center O
(which is the center of buoyancy). The displacement determines the
particle orientation, defined by the unit vector
${\bm p}$ directed opposite to the center of mass.
The direction ${\bm p}$ is determined by the balance between the different
torques acting on the particle.  Because of particle asymmetry, an external
force ${\bm f}$, such as gravity $m {\bm g}$, exerts a torque
red${\bm T}_f=-h {\bm p} \times {\bm f}$.  In addition, the
particle immersed in a fluid experiences a viscous torque ${\bm T}_v=8 \pi \nu
\rho r^3 ({\bm \omega}/2-{\bm \Omega})$, where ${\bm \omega}={\bm \nabla}
\times {\bm u}$ is the fluid vorticity, $\nu$ and $\rho$ are the fluid
kinematic viscosity and density, respectively, and ${\bm \Omega}$ is the
angular velocity of the sphere.  If the particle Reynolds number is very small,
we can assume creeping flow conditions around the sphere, which impose
equilibrium between the external forces and the viscous ones, in this
particular case zero total torque ${\bm T}_f + {\bm T}_v=0$.
From the solid body rotation formula $\dot{\bm p}={\bm p}\times{\bm \Omega}$
and ${\bm p} \times ({\bm
  T}_f + {\bm T}_v)=0$ we end with the following equation for the
orientation \cite{pedley_arfm92}
\begin{equation}
{d{\bm p} \over dt} = -{1 \over 2 v_0} \left[{\bm A} - ({\bm A} \cdot {\bm p})
{\bm p} \right] + {1 \over 2} {\bm \omega} \times {\bm p} \,,
\label{eq:2}
\end{equation}
where ${\bm A}$ denotes the total acceleration 
(due to the flow and gravity) on the sphere and the constant $v_0=3 \nu/h$,
having the dimension of a velocity, weighs the contribution of external forces
to particle orientation. 
We remark that in the case of axisymmetric non-spherical particles an additional
term is present in (\ref{eq:2}) \cite{pedley_arfm92,Vincenzi2012,Parsa2012}.

Equations (\ref{eq:1}) and (\ref{eq:2}) are valid in
  the limit of small, neutrally buoyant particles. If inertia is taken
  into account, particle motion is described by integro-differential
  equations containing added mass and history effects (see, e.g.,
  Ref.~\cite{Maxey}). No such effects will be considered here,
  consistently with our approximations. In a fluid at rest, the
only external force entering Eq.~(\ref{eq:2}) is gravity, so that
${\bm A}={\bm g}=(0,0,-g)$. The result is that particles orient
upwards in a time $\mathcal{O}(v_0/g)$. Such phenomenon is well known
in bio-fluid-dynamics, and is at the basis of the ability of some
bottom-heavy phytoplankters to swim towards the sea surface (a
phenomenon dubbed \textit{negative gravitaxis} \cite{pedley_arfm92}),
maximizing the exposition to light and thus the photosynthetic
activity.

In a turbulent flow, advected particles are subject to intense
accelerations, so that, locally, gravity must be corrected due to
inertial forces.  The total acceleration $\bm A$ acting on the
particle is thus given by ${\bm A}={\bf g}-{\bm a}$, where ${\bm
  a}=d{\bm u}/dt$ is the fluid acceleration at the particle position.
 Again, the assumption that the acceleration of the
  particle is equal to that of the fluid implies particles smaller
  than few $\eta$.  Numerical \cite{Calzavarini2009,Homann2010} and
  experimental \cite{Voth2002, Qureshi2007} investigations showed that
  particles larger than $\eta$ sense accelerations smaller than
  tracers. If one restricts to diameters up to $4 \eta$ the error on
  the rms value should be less than $20\%$.  There is indication that
  such larger particles can accurately be described by including
  Faxen's corrections in the equation of motion \cite{Homann2010,
    Calzavarini2009}.  

In the following, for the sake of simplicity, we assume that fluid
acceleration is smaller than gravity, i.e. $g\gg a_{\rm
  rms}$. This is true for flows at moderate Reynolds
  numbers only \cite{LaPorta2001}, but this assumption greatly
  simplifies the analysis of (\ref{eq:2}).  Formally, we can write
  ${\bm A}={\bm g} - \epsilon {\bm a}$ in Eq.~(\ref{eq:2}), with
  $\epsilon$ a small non-dimensional number.  Further we consider the
  limit of fast reorientation, which amounts to
    requiring that the reorientation time $v_0/g$ is smaller than the
    Kolmogorov time $\tau_\eta$. If one estimates $a_{rms}\sim
    \epsilon^{3/4}/\nu^{1/4}$ and assumes $h\sim \eta$ in the
    definition of $v_0$, fast orientation consistently implies $g\gg
    a_{\rm rms}$.

When the vortical term ${\bm \omega} \times {\bm p}$ 
is small, i.e. when $v_0 \omega_{rms} \ll a_{rms}$, 
 Eq.~(\ref{eq:2}) reduces to $\bm A=(\bm
A\cdot \bm p)\bm p$, which explicitly reads

\begin{eqnarray}
\epsilon a_x &=& (\epsilon {\bm a} \cdot {\bm p} + g p_z) p_x \nonumber \\
\epsilon a_y &=& (\epsilon {\bm a} \cdot {\bm p} + g p_z) p_y \label{eq:3}\\
\epsilon a_z +g &=& (\epsilon {\bm a} \cdot {\bm p} + g p_z) p_z\,. \nonumber 
\end{eqnarray}
From (\ref{eq:3}) one can see that the orientation vector must have
the form ${\bm p}=(\epsilon q_x,\epsilon q_y, 1)$, indeed as
$p^2=1$ the correction to $p_z$ will be $\mathcal{O}(\epsilon^2)$ so
that we can neglect it at this level. Plugging the expression for $\bm
p$ in Eq.~(\ref{eq:3}), at $\mathcal{O}(\epsilon)$ we obtain ${\bm
  q}=(a_x/g,a_y/g,0)$. In conclusion, measuring the orientation of the
particle with respect to the vertical we can measure two components of
the fluid acceleration
\begin{equation}
{\bm p}=\left(\frac{a_x}{g}, \frac{a_y}{g}, 1 \right)
\label{eq:4}
\end{equation}
This result is valid under the assumption that $a_{rms} \ll g$ and,
therefore, in general for not too high Re. This condition together
with the smallness of the vorticity term implies fast orientation
($v_0/g\ll\tau_\eta$).  As a consequence, if $v_0$ is too large, the
first effect we expect is that vorticity becomes relevant, with an
increase of the tilting angle, so that using (\ref{eq:4}) could
  lead to an over-estimate of accelerations. In more general
conditions, it is in principle still possible to use (\ref{eq:2}) to
gather information
on the acceleration statistics but this requires less direct
procedures, which we will not consider in this preliminary study.

%%%%%%%%%%%%%%%%%%%%%%%%%%%%%%%%%%%%%%%%%%%%%%%%%%%%%%%%%%%%%%%%%%%%%%
\section{Numerical simulations of geotropic tracers in turbulence}
\label{sec3}

In this section we illustrate the behavior of geotropic particles in
realistic turbulent flows and explore the range of validity of the
result (\ref{eq:4}) by means of DNS of
the dynamics of geotropic tracers together with the Navier-Stokes
equations for an incompressible flow. 
Trajectories (up to $2 \times 10^5$) are stored together with $\bm a$
and $\bm \omega$ in statistically stationary conditions.
Equation~(\ref{eq:2}) is then integrated starting from
random orientations and for different values of $v_0$.  After an initial transient of the order of
$v_0/g$, during which particles forget their initial orientation, we can
compare the acceleration ${\bm a}$ with the prediction of
Eq.~(\ref{eq:4}).

We have performed simulations of homogeneous-isotropic turbulence by
means of a parallel pseudo-spectral code in a cubic box with periodic
boundary conditions at ${\rm Re}_{\lambda} \simeq 200$ with resolution
$512^3$. Statistical stationarity was maintained via a Gaussian,
delta-correlated in time, random forcing at small wave-numbers.
Eq.~(\ref{eq:1}) is integrated evaluating the velocity at
particle position by means of trilinear interpolation. 
  Moreover, we have also exploited a database
\cite{database} of previously simulated Lagrangian trajectories at
resolution $2048^3$ and ${\rm Re}_{\lambda} \simeq 400$, for which
acceleration and vorticity were available. Equation (\ref{eq:2}) was integrated
using a second-order Adams-Bashforth scheme. In order to get physical
relevance from the DNS we rescale space and time with dimensional
values.  This is easily done by matching the Kolmogorov scale and time
with experimental values at similar ${\rm Re}_{\lambda}$, as shown in
Table~\ref{table1}. We remark that this rescaling is not unique as
${\rm Re}_{\lambda}$ fixes a ratio of scales (and times) and not an
absolute scale. This point is crucial as the parameter $v_0$ is
limited by the size of the particle and $g$ is obviously fixed.  In
the following we will use laboratory experiments with an integral scale
of the order of few $cm$ for rescaling our simulations to physical
values \cite{Voth2002,Klein2013}. 

%------------------------------------------------------------------------
\begin{table}[h]
\begin{tabular}{c|c|c|c|c|c|c|c}
${\rm Re}_{\lambda}$ & $\eta$ & $\tau_{\eta}$  & $u_{\rm rms}$ 
& $\epsilon$ & $L$ & $a_{\rm rms}$ & $T$ \\
 & (${\rm m}$) & (${\rm s}$) &(${\rm m} {\rm s}^{-1}$) 
& (${\rm m}^2 {\rm s}^{-3}$) & (${\rm m}$) & (${\rm m} {\rm s}^{-2}$) & $ ({\rm s})$ \\
\hline 
$200$ & $191 \times 10^{-6}$ & $37 \times 10^{-3}$ & $0.037$ & 
$7.1 \times 10^{-4}$ & $0.07$ & $0.24$ & $8.22$ \\
$400$ & $98 \times 10^{-6}$ & $9.5 \times 10^{-3}$ & $0.097$ & 
$9.1 \times 10^{-3}$ & $0.10$ & $1.5$ & $2.58$
\end{tabular}
\caption{Parameters of the simulations made dimensional on the
basis of laboratory experiments at similar Reynolds 
numbers \cite{Voth2002,Klein2013}.
$\eta=(\nu^3/\epsilon)^{1/4}$ is the Kolmogorov scale,
$\tau_{\eta}=(\nu/\epsilon)^{1/2}$ the Kolmogorov time,
$u_{\rm rms}$ is the 
root mean square of the velocity, $\epsilon$ is the energy dissipation per
unit mass, $L=u'^3/\epsilon$ is the integral length scale, 
$a_{\rm rms}$ is the root mean square acceleration.
$T$ is the integration time. For both simulations $g=9.8 {\rm m} {\rm s}^{-2}$.}
\label{table1}
\end{table}
%------------------------------------------------------------------------

%------------------------------------------------------------------------
\begin{figure}[htb!]
\centering
\includegraphics[width=14cm]{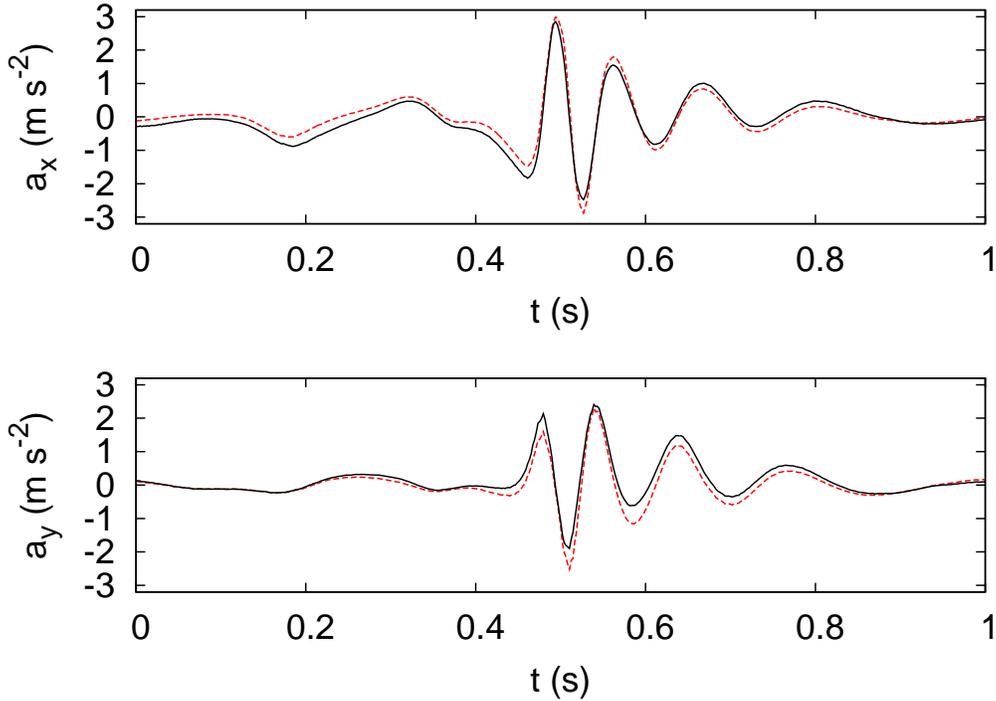}
\caption{$x$ component (top) and $y$ component (bottom) of the
  acceleration of one particle computed from numerical simulations at
  ${\rm Re}_{\lambda}=200$ (black line) together with the acceleration
  estimated from the $x$ and $y$ component of the orientation vector,
  i.e. $a_x=g p_x$ $a_y=g p_y$ see (\ref{eq:4}), of a geotropic
  particle with $v_0=6 {\rm mm}~{\rm s}^{-1}$ corresponding to a displacement
  $h=0.5 {\rm mm}$.}
\label{fig2}

\end{figure}
%------------------------------------------------------------------------

Fig.\ref{fig2} shows an example of time series of the two components
of the acceleration $a_x$ and $a_y$ obtained following a Lagrangian
tracer in the flow at ${\rm Re}_{\lambda}=200$.  The initial condition
of the orientation is along the $z$ axis, ${\bm p}(0)=(0,0,1)$.  The
dashed red line represents the acceleration obtained
according to (\ref{eq:4}) from the $x$ component of the orientation
vector, $a_x=g p_x$ of a particle with $v_0 \simeq 0.006 {\rm m} {\rm s}^{-1}$.
The corresponding relaxation time under gravity is $\tau=v_0/g \simeq
6\times 10^{-4} s$.  In this case $a_{rms} \ll g$ and, therefore, the
estimation (\ref{eq:4}) is fully justified and indeed the acceleration
is reproduced quite accurately.

%------------------------------------------------------------------------
\begin{figure}[htb!]
\centering
\includegraphics[width=14cm]{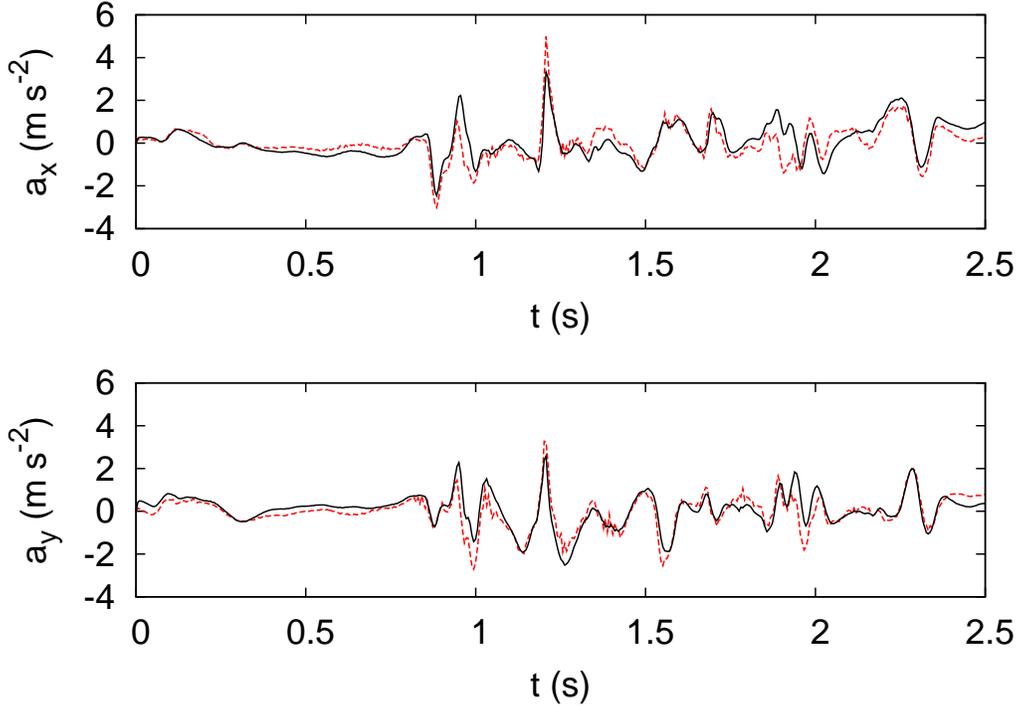}
\caption{The same of Fig.\ref{fig2} for a geotropic trajectory
  with $h=0.2 {\rm mm}$ and $v_0=15 {\rm mm}~{\rm s}^{-1}$ in a
  turbulent flow at ${\rm Re}_{\lambda}=400$.
\label{fig3}}
\end{figure}
%------------------------------------------------------------------------

In Fig.\ref{fig3} we show an example for a trajectory in a turbulent
flow at ${\rm Re}_{\lambda}=400$. Although the rms of acceleration
$a_{rms}\simeq \epsilon^{3/4} \nu^{-1/4}$ is smaller than $g$,
particles experience fluctuations comparable to, or even larger than $g$, 
where the assumptions leading to (\ref{eq:4}) are not applicable. These large
fluctuations of Lagrangian acceleration are typical in turbulence and
physically correspond to event of trapping of tracers in small scale vortices
\cite{LaPorta2001,Biferale2004,Biferale2005}. As shown in Fig.\ref{fig3},
during these events the orientation vector ${\bm p}$ is unable to fully follow
the acceleration fluctuation, which results to be slightly underestimated.
%------------------------------------------------------------------------
\begin{figure}[htb!]
\centering
\includegraphics[width=0.48\textwidth]{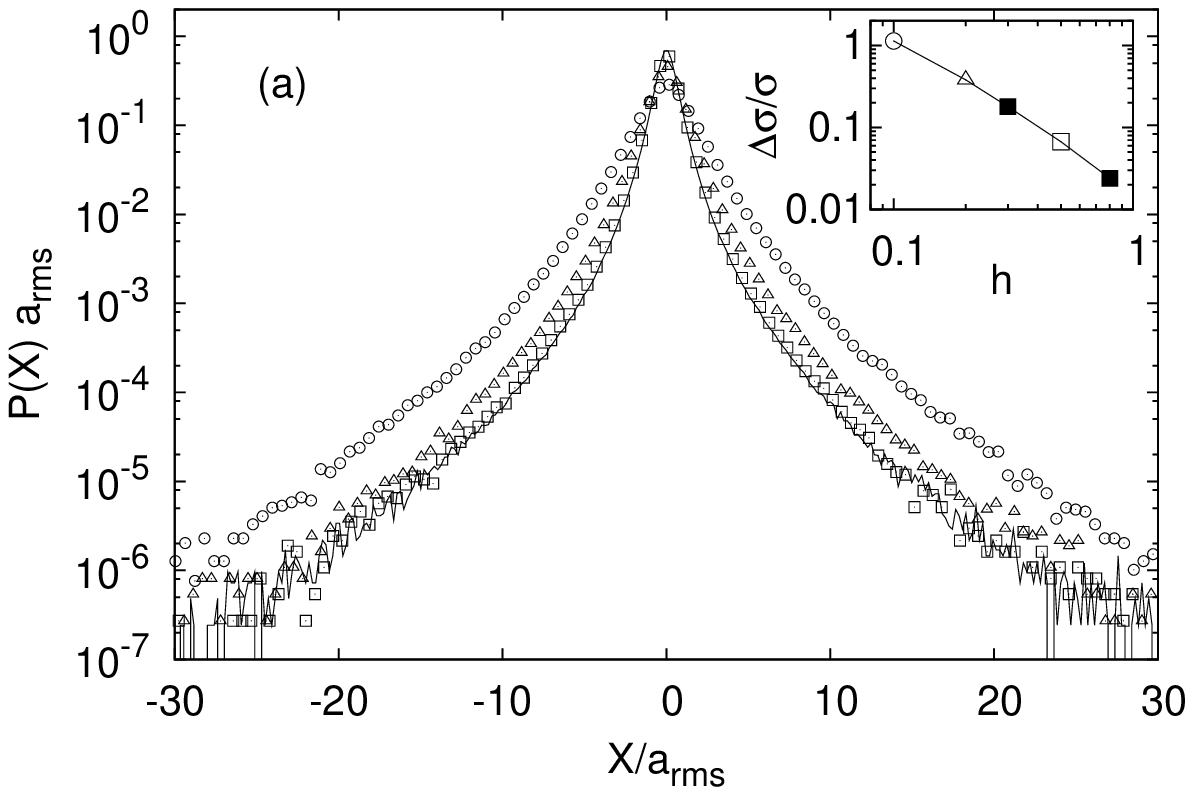}
\includegraphics[width=0.48\textwidth]{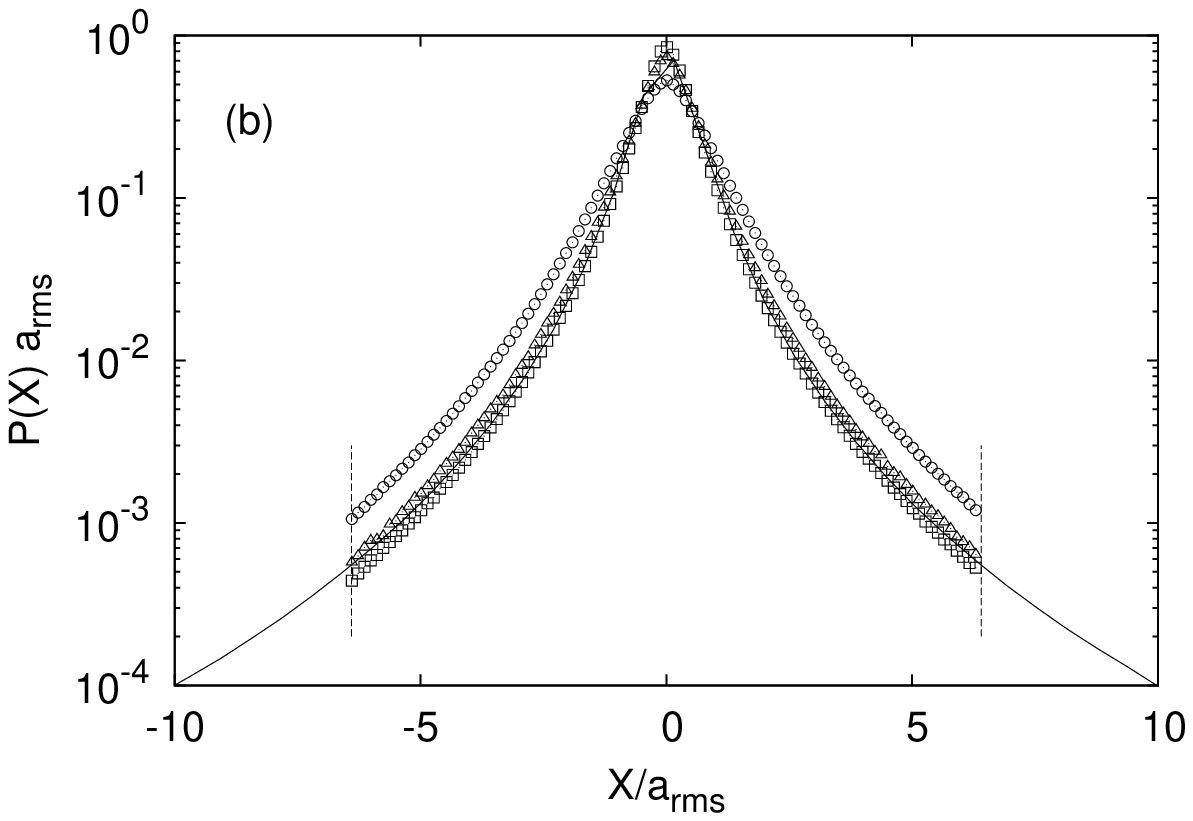}
\caption{PDFs of acceleration in one horizontal direction for ${\rm
    Re}_\lambda=200$ (a) and ${\rm Re}_\lambda=400$ (b). Estimates
  obtained according to (\ref{eq:2}) ($X=gp_x$, symbols) are compared
  with fluid acceleration ($X=a_x$, line).  Three values of particle
  bias were used, which rescaled on experimental values correspond to
  $h=0.1 {\rm mm}$ (circles), $0.2 {\rm mm}$ (triangles) and $0.5 {\rm
    mm}$ (squares). By comparison, it is evident that the intermediate
  value gives a good estimate at higher ${\rm Re}_\lambda$ but is not
  satisfying at the lower one(see text). In (b) the value of $g$
  (vertical lines) marks the upper cutoff for measurable
  accelerations.  In the inset of  (a): relative error on the
  estimate of $\sigma=\sqrt{\langle a_x^2\rangle}$ as a function of
  $h$, for ${\rm Re}_\lambda=200$, $\Delta\sigma=g\sqrt{\langle
    p_x^2\rangle}-\sigma$.\label{fig4}}
\end{figure}
%------------------------------------------------------------------------

On a more quantitative level, Fig.\ref{fig4} shows the probability
density function (PDF) of acceleration compared with the estimation
obtained via (\ref{eq:4}).  For each value of ${\rm Re}_\lambda$ considered
we simulate the results of three hypothetical experiments, with
particles of different sizes. As discussed above, geotropic
orientation is expected to be a good proxy for acceleration only in
the limit of small $v_0$, i.e. for fast orientation. As apparent from
both panels in Fig.\ref{fig4}, when a large enough displacement $h$
is considered, the statistics of acceleration is reproduced remarkably
well by particle orientation. However, this is not the case if less
biased particles are considered. This clearly implies a {\em lower}
limit in the size of particles used, a factor that must be taken into
account in the design of possible experiments. As mentioned above,
vorticity can be neglected only if $v_0\omega_{rms}/a_{rms}\sim
v_0/\delta u_\eta<1$. By applying the definition of the Kolmogorov
scale $\delta u_\eta \eta/\nu=1$ and that of $v_0$, the constraint
reduces (a part from order-one coefficients) to $\eta\lesssim
h$. This inequality can pose a problem both for the validity of (\ref{eq:1})
and the actual statistics seen by the particle. Both points will be discussed
in the final section. As for now we will just consider this condition in the
framework of our model, assuming that the corrections are small as long as
the particle size is of the same order as $\eta$. 

If (as in our case) one considers a set of experiments all using
water and with comparable integral scales, an increase in Re
corresponds to a smaller viscous scale, thus decreasing the minimum
particle size required to reconstruct acceleration. As an example of
this we considered the case of particles with $h=0.2 {\rm mm}$. As
evident from Fig.\ref{fig4} using Eq.~(\ref{eq:4}) on statistics obtained
with such particles would lead to an overestimate of larger
accelerations at ${\rm Re}_\lambda=200$ (triangles in Fig.\ref{fig4}a)
while they would be acceptable candidates at ${\rm Re}_\lambda=400$
(triangles in Fig.\ref{fig4}b). 
However, the largest acceleration that can be measured by means of (\ref{eq:2})
is $g$. For experiments at higher Re where very large accelerations are
present, this introduces a cut-off in the estimated accelerations. As
evident from the results at ${\rm Re}_\lambda=400$ the core of the PDF
is approximately correct, even if values above $0.5\div 0.7 g$ are
underrepresented. We stress that the simulations exhibited
accelerations up to 80 $a_{rms}$ (not shown for graphical reasons),
while $g\approx 6.3 a_{rms}$ if rescaled over the experimental
parameters. Analysis of the variance of acceleration performed for ${\rm
Re}_\lambda=200$ (inset of Fig.\ref{fig4}a) reveals that  the second moment of
the distribution is correctly recovered asymptotically in $h/\eta$. The same
cannot be verified at ${\rm Re}_\lambda=400$, since the cut-off at $g$ prevents
convergence of the second moment of estimated accelerations.

%------------------------------------------------------------------------
\begin{figure}[htb!]
\centering
\includegraphics[width=0.7\textwidth]{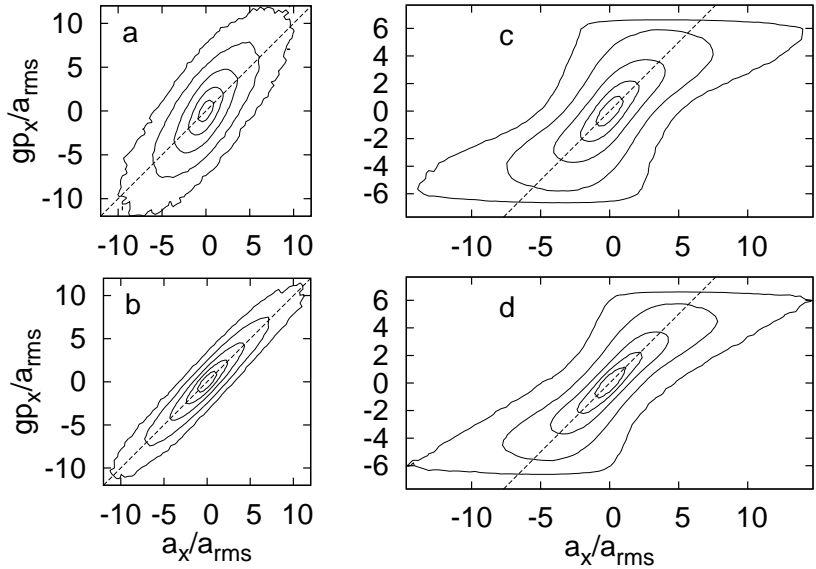}
\caption{Joint PDF of acceleration and estimated
    acceleration, for ${\rm Re}_\lambda=200$ (a,b) and ${\rm
      Re}_\lambda=400$ (c,d). Each panel refers to a different value
    of the displacement, $0.2 {\rm mm}$ (a,c) and $0.5 {\rm mm}$
    (b,d). Contour levels are set a factor 10 apart starting from
    $10^{-1}$ (at the centre) down. The straight line marks $gp_x=a_x$
    for reference. While the tendency is generally that of
    overestimating large accelerations (appearing as a clockwise tilt
    of the level sets), stronger "clockwise" lobes appears for the
    larger displacement at ${\rm Re}_\lambda$ (d), compatible with the
    lower tails in the corresponding PDF of Fig.\ref{fig4}. Note that
    the strong deformation of the PDF in (c) and (d) is due to the
    cut-off $gp_x$.}\label{fig5}
\end{figure}
%------------------------------------------------------------------------

%------------------------------------------------------------------------
\begin{figure}[htb!]
\centering
\includegraphics[width=1\textwidth]{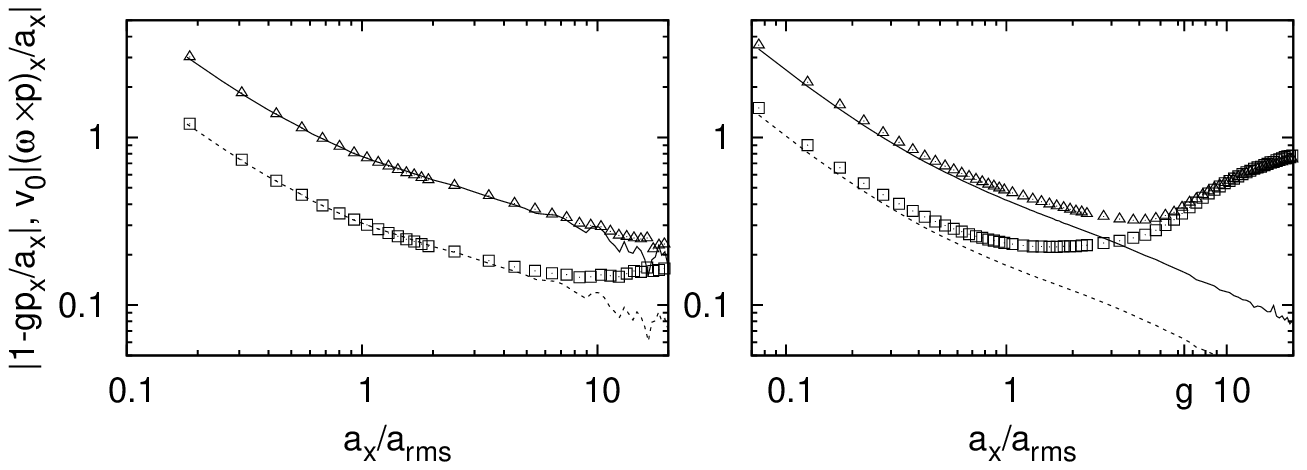}
\caption{Relative error on the estimate of one component of
  acceleration by (\ref{eq:4}). The average $\langle |1-gp_x/a_x|;
  a_x\rangle$ conditioned on the local value of $a_x$ (symbols) is
  compared with the contribution due to the vorticity term in
  (\ref{eq:2}) $\langle v_0|\omega_y p_z-\omega_zp_y|; a_x\rangle$
  (lines). For ${\rm Re}_\lambda=200$ (left), the latter clearly
  constitutes the main contribution to the error. At ${\rm
    Re}_\lambda=400$ (right), the estimate of larger accelerations is
  clearly affected by the finite value of $g$. Data refer to
  $h=0.2{\rm mm}$ (triangles, solid line) and $h=0.5{\rm mm}$ (squares,
  dotted line).}
 \label{fig6}
\end{figure}
%------------------------------------------------------------------------

 In order to further investigate the errors on the
  estimate of the acceleration, we consider the joint distribution
  $P(a_i,gp_i)$ (with $i=x,y$) of each acceleration component and
  its estimate. As show in Fig.\ref{fig5} such distributions confirm a
  tendency of smaller particles to overestimate accelerations. Only
  for ${\rm Re}_\lambda=400$ the largest particles underestimate
  accelerations, as can be seen by the low tails of the corresponding
  PDF in Fig.\ref{fig4} and by a slight asymmetry of $P(a_x,gp_x)$
  towards quadrants in which $|gp_x|<|a_x|$.  The strongly
  intermittent nature of both acceleration and vorticity suggests to
  investigate in more detail how accurately accelerations of different
  magnitude can be estimated via (\ref{eq:4}). The conditional average
  $\langle|1-g p_x/a_x|;a_x\rangle$ is shown in Fig.\ref{fig6} for both values
  of ${\rm Re}_\lambda$. Let us first consider the curves at
  ${\rm Re}_\lambda=200$. It is evident that larger particles (i.e. with
  faster reorientation time) provide better estimates: the largest
  particles, with $h=0.5 {\rm mm}$ give a minimum relative error of around
  0.2. Through most of the observed range, the relative error is
  smaller for larger accelerations, because the effect of vorticity
  decreases accordingly. Indeed, the same figure also compares the
  relative error with $v_0(\omega\times\bm p)$, showing that, for all
  but the largest accelerations, the error in the estimate comes from
  the vorticity term in (\ref{eq:2}), consistently with the assumption
  of fast orientation. For accelerations larger than $\sim0.1g$ the
  effect of finite gravity causes deviations from this behaviour and
  eventually an increase of the relative error, as expected. This
  effect is less evident for smaller particles, most likely because
  they tend to overestimate the acceleration while the finite gravity
  effect leads to an underestimate so that there is a compensation
  between the two opposite effects. The right panel shows that the
  effect of finite gravity is much larger for ${\rm Re}_\lambda=400$,
  as expected. However, one should note that the vorticity term would
give with the same particles a smaller error in this second case than for ${\rm
Re}_\lambda=200$. Indeed, by estimating the error due to vorticity as
$v_0/u_\eta$ one would get a value about 1.9 smaller for the higher ${\rm
Re}_\lambda$, compatible within $10\%$ with the numerical results around $a\sim a_{\rm rms}$.  We stress that this observation is not valid in general: it is a consequence of the fact that, in our case, the flow at higher Re has a larger effective integral scale and a larger $u_\eta$.

%%%%%%%%%%%%%%%%%%%%%%%%%%%%%%%%%%%%%%%%%%%%%%%%%%%%%%%%%%%%%%%%%%%%%%
\section{Conclusion and discussion}
\label{sec4}
Summarizing our numerical results, it appears that the orientation of biased particles could be a viable proxy for fluid acceleration, at least at moderate Re. It is clearly important to establish a way to estimate the proper particle size based on the parameters of the turbulent flow to be examined.

Although particle size does not directly enter the
  model equations, the offset $h$ clearly puts a lower limit on
particle radius. This point must be carefully considered. Particles
should be sufficiently biased to ensure dominance of acceleration over
rotation due to vorticity, but too large particles would not
obey the assumptions leading to (\ref{eq:1}) and (\ref{eq:2}).

On the other hand, experimental limitations should be considered. 
In order to use (\ref{eq:4}) to directly measure fluid acceleration
one has to measure the tilt angle of a geotropic particle transported
by the flow. One possibility is to use small spherical particles
with the upper and lower hemisphere of different colors as in the example
of Fig.\ref{fig1}, a simpler version of the technique used in
\cite{Zimmermann2011}. By measuring the angle $\theta$ of the particle
``equator'' with respect to the horizontal plane one has $p_x=\sin \theta$.

A precise determination of $\theta$ requires sufficient resolution of 
the particle pattern and therefore not too small particles. On the other
hand, because the measure is instantaneous, there is no need to follow
the particles. Therefore, the camera could be placed to zoom a small region
of the fluid only, and to acquire data when a particle comes in that region.

Let us finally comment about possible corrections to the described
behavior, for the two cases of particles too small or too large. If
the offset $h$ is too small, the vorticity term in (\ref{eq:2}) is no
longer negligible. As a consequence, reconstruction of acceleration
would require independent information on vorticity, so that a more
complex method would be required.  Furthermore, fast orientation is at
the basis of (\ref{eq:4}) which allows one to avoid particle tracking,
and would be important to follow high frequency fluctuations
accurately. In the case of too large particles, the creeping flow
assumption would be inaccurate.  Nonetheless, the tilting angle would
still provide information on the fluid acceleration but equation
(\ref{eq:2}) has to be modified to take inertial terms into account.

A further aspect that should be taken into account is finite size
effects on particle trajectory.  In general one expects that particles
larger than the Kolmogorov scale deviate from fluid
trajectories. However there is evidence that acceleration statistics
are not strongly influenced by particle size
\cite{Qureshi2007,Homann2010, Calzavarini2009}. Numerical and
experimental results suggest that addition of the so called Faxen
terms in the equation for particle trajectory can account for the main
deviations, providing a method to estimate the related
errors\cite{Calzavarini2009, Homann2010}. Such corrections should
become relevant when the radius of the particle is larger than
$\eta\sqrt{{\rm Re}_\lambda}$, which for experiments comparable to the
ones we considered would give $O(10)\eta$ \cite{Calzavarini2009} thus
allowing for some range of sizes to explore.

Given the above constraints we can conclude that the
  proposed method would be reasonably accurate for typical
  experimental settings at moderate Reynolds numbers or when large
  Reynolds number are achieved thanks to a large integral scale.  In
spite of the above discussed limitations, we think that the idea of
exploiting biased particles to measure acceleration without tracking
may be interesting especially if technology can be pushed to the
possibility to measure the tilting angle of many particles at the same
time, allowing for the reconstruction of the spatial field of
accelerations.

%------------------------------------------------------------------------


\begin{thebibliography}{18}
\providecommand{\natexlab}[1]{#1}

\bibitem[1]{Toschi2009}
F. Toschi, and E. Bodenschatz, {\itshape Lagrangian properties of particles in
  turbulence}, Annual Review of Fluid Mechanics 41 (2009), pp. 375--404.

\bibitem[2]{Arneodo2008}
A. Arn{\`e}odo et~al., {\itshape Universal intermittent properties of particle
  trajectories in highly turbulent flows}, Physical Review Letters 100 (2008),
  p. 254504.

\bibitem[3]{Biferale2004}
L. Biferale, G. Boffetta, A. Celani, B. Devenish, A. Lanotte, and F. Toschi,
  {\itshape Multifractal statistics of Lagrangian velocity and acceleration in
  turbulence}, Physical review letters 93 (2004), p. 64502.

\bibitem[4]{Zimmermann2011}
R. Zimmermann, Y. Gasteuil, M. Bourgoin, R. Volk, A. Pumir, and J.F. Pinton,
  {\itshape Rotational Intermittency and Turbulence Induced Lift Experienced by
  Large Particles in a Turbulent Flow}, Phys. Rev. Lett. 106 (2011), p.
  154501.


\bibitem[5]{Zimmermann2011a}
R. Zimmermann, Y. Gasteuil, M. Bourgoin, R. Volk, A. Pumir, and J.F. Pinton,
  {\itshape Tracking the dynamics of translation and absolute orientation of a
  sphere in a turbulent flow}, Review of Scientific Instruments 82 (2011), p.
  033906.

\bibitem[6]{Zimmermann2012}
R. Zimmermann, L. Fiabane, Y. Gasteuil, R. Volk, and J. Pinton, {\itshape
  Measuring Lagrangian accelerations using an instrumented particle},  (2012),
  arXiv preprint arXiv:1206.1617.

%\bibitem[7]{Klein2012}
%S. Klein, M. Gibert, A. B\'erut, and E. Bodenschatz, {\itshape Simultaneous 3D
  %measurement of the translation and rotation of finite size particles and the
  %flow field in a fully developed turbulent water flow},  (2012),  arXiv
  %preprint arXiv:1205.2181.

\bibitem[7]{Klein2013}
S. Klein, M. Gibert, A. B\'erut, and E. Bodenschatz, {\itshape Simultaneous 3D
  measurement of the translation and rotation of finite size particles and the
  flow field in a fully developed turbulent water flow}, Measurement Science and Technology 24 (2013), p. 024006

\bibitem[8]{Vincenzi2012}
D. Vincenzi, {\itshape Orientation of non-spherical particles in an
  axisymmetric random flow},  (2012),  arXiv preprint arXiv:1206.0945.

%\bibitem[9]{Parsa2012}
%S. Parsa, E. Calzavarini, F. Toschi, and G. Voth, {\itshape Rotation rate of
  %rods in turbulent fluid flow},  (2012),  arXiv preprint arXiv:1205.0219.

\bibitem[9]{Parsa2012}
S. Parsa, E. Calzavarini, F. Toschi, and G. Voth, {\itshape Rotation rate of
  rods in turbulent fluid flow} Physical Review Letters 109 (2012), p. 134501


\bibitem[10]{pedley_arfm92}
T.J. Pedley, and J.O. Kessler, {\itshape Hydrodynamic Phenomena in Suspensions
  of Swimming Microorganisms}, Annual Review of Fluid Mechanics 24 (1992), pp.
  313--358.

\bibitem[11]{Maxey}
M.R. Maxey, and J.J. Riley, {\itshape Equation of motion for a small rigid
  sphere in a nonuniform flow}, Physics of Fluids 26 (1983), p. 883.

\bibitem[12]{Calzavarini2009}
E. Calzavarini, R. Volk, M. Bourgoin, E. Leveque, J. Pinton, and F. Toschi,
  {\itshape Acceleration statistics of finite-sized particles in turbulent
  flow: the role of Fax{\'e}n forces}, Journal of Fluid Mechanics 630 (2009),
  p. 179.

\bibitem[13]{Homann2010}
H. Homann, and J. Bec, {\itshape Finite-size effects in the dynamics of
  neutrally buoyant particles in turbulent flow}, Journal of Fluid Mechanics
  651 (2010), p.~81.

\bibitem[14]{Voth2002}
G. Voth, A. la  Porta, A. Crawford, J. Alexander, and E. Bodenschatz, {\itshape
  Measurement of particle accelerations in fully developed turbulence}, Journal
  of Fluid Mechanics 469 (2002), pp. 121--160.

\bibitem[15]{Qureshi2007}
N.M. Qureshi, M. Bourgoin, C. Baudet, A. Cartellier, and Y. Gagne, {\itshape
  Turbulent Transport of Material Particles: An Experimental Study of Finite
  Size Effects}, Phys. Rev. Lett. 99 (2007), p. 184502.

\bibitem[16]{LaPorta2001}
A. La~Porta, G. Voth, A. Crawford, J. Alexander, and E. Bodenschatz, {\itshape
  Fluid particle accelerations in fully developed turbulence}, Nature 409
  (2001), pp. 1017--1019.

\bibitem[17]{database}
F. Toschi, L. Biferale, M. Cencini, E. Calzavarini, A. Lanotte, and J. Bec,
  {\itshape Heavy particles in turbulent flows RM-2007-GRAD-2048.St0.
  iCFDdatabase. Dataset.},  (2011),
  http://dx.doi.org/10.4121/uuid:a64319d5-1735-4bf1-944b-8e9187e4b9d6.

\bibitem[18]{Biferale2005}
L. Biferale, G. Boffetta, A. Celani, A. Lanotte, and F. Toschi, {\itshape
  Particle trapping in three-dimensional fully developed turbulence}, Physics
  of Fluids 17 (2005), p. 021701.

\end{thebibliography}
\end{document}